\documentclass[twocolumn]{aastex62}
\pdfoutput=1

\usepackage{subfigure}
\usepackage{textcomp}
\usepackage{amsmath}
\usepackage{natbib}
\usepackage{xcolor}
\usepackage{fleqn}
\begin{document}

\title{Laboratory Impact Splash Experiments to Simulate Asteroid Surfaces}

\email{tabea.bogdan@uni-due.de}

\author{Tabea Bogdan}
\affil{University of Duisburg-Essen, Faculty of Physics, Lotharstr. 1-21, 47057 Duisburg, Germany}

\author{Jonathan E. Kollmer}
\affil{University of Duisburg-Essen, Faculty of Physics, Lotharstr. 1-21, 47057 Duisburg, Germany}

\author{Jens Teiser}
\affil{University of Duisburg-Essen, Faculty of Physics, Lotharstr. 1-21, 47057 Duisburg, Germany}

\author{Maximilian Kruss}
\affil{University of Duisburg-Essen, Faculty of Physics, Lotharstr. 1-21, 47057 Duisburg, Germany}

\author{Gerhard Wurm}
\affil{University of Duisburg-Essen, Faculty of Physics, Lotharstr. 1-21, 47057 Duisburg, Germany}

\begin{abstract}
Granular material that is bound by the low gravity of a small asteroid is mobilized by slow velocity impacts. These splashes generated by impacts might play an important role in sculpting the asteroid's surface. In laboratory experiments we characterize the ejecta generated by spherical 150\,$\rm \mu m$ diameter basalt grains impacting a granular bed at 0.8\,m/s.  We find that such an impact typically leads to less than 10 particles being ejected from the granular bed, with typical ejecta trajectories rising to less than one particle diameter above the surface. That is, the observed impacts are highly dissipative and only a small fraction of the impact energy is imparted onto the ejecta. While the impactor itself still rebounds, it typically slows down significantly to an average of about 20\,\% of its impact velocity. Scaled to asteroids, impactor and ejecta generated from impacts of sand sized grains are not able to spread over the asteroid's surface but will stay close to the impact site. Therefore these highly inelastic impacts into soft granular beds efficiently trap grains, in contrast to more elastic impacts on bare, rocky surfaces confirming suggestions by \citet{Shinbrot2017}. This is also in agreement to observed features on asteroids as this topological elasticity bias suggests that redistribution of grains leads to a size segregation. 
\end{abstract}


\section{Introduction}

The surface of small rubble pile asteroids often shows regions which are dominated by different grain sizes. There are bare rocky regions contrasting pond or sea like granular beds of small grains. Examples are the small asteroids Itokawa \citep{Saito2006} or Eros \citep{Veverka2000}.  
With several km in size, the gravitational acceleration is only on the order of $ 10^{-4} \dots 10^{-5} \rm\,m/s^2$ and the escape velocity can be as small as a few cm/s. 
There are often more {\it ponds} of small grains at different locations spread across the surface and the small grains do not necessarily all gather at the global gravitational minimum. Granular beds are found at similar elevations as rocks next to it \citep{Saito2006}. Downhill flow surely is a part of the explanation \citep{susorney2019}. However, additional mechanisms might be at work here. 
It is, e.g., not obvious why bare, rocky regions are so clean of small sand and dust sized grains. At the observed low gravity levels, even sand sized particles with low cohesion could easily stick to rocks.

Several mechanisms have been discussed for shaping these surfaces. Grains charge by electromagnetic and particle radiation (UV, solar wind, cosmic radiation). Electrostatic forces can eject and redistribute particles \citep{Colwell2005}. Laboratory experiments by \cite{Wang2010} show that particles indeed jump off a dust bed if subject to UV or a directed electron beam.
Recently, \cite{demirci2019} studied the erosion of weakly cohesive
matter under reduced gravity at low pressure. This work and especially earlier works by \citet{Thomas2015} and \citet{Jia2017} suggest that segregation might have occurred while asteroids were still in the protoplanetary disk. This has not been worked out in detail though and it is unknown how primordial the surfaces of asteroids actually are. At least some areas on asteroids show cratered surfaces suggesting continuous impact shaping beyond the protoplanetary disk phase \citep{Sugita2019}.

A different approach is followed by \citet{Guettler2013} which assume a brazil nut effect to be active. They carried out microgravity experiments to study the gravity dependence of the brazil nut effect. 
However, the details are highly dependent on the environmental conditions \citep{Schroeter2006}. Furthermore, if a body with such a low gravity as for example Itokawa is shaken so violently that significant parts of it are in motion, one can assume that it will eject a significant fraction of grains. The fate of these ejecta escaping or re-impacting the asteroid will be important then. 

This connects to the mechanism studied here. Impacts of grains onto a granular bed and a bare solid surface have been tested in first experiments by \citet{Shinbrot2017}. They find that smaller particles will concentrate on an already existing granular bed, essentially leaving the bare surface as it is. The idea is simply that the collision with a granular bed is much more inelastic than the collision with a solid target. Rebounding particles are then less likely to escape a granular bed (as compared to impacts on a solid surface) and thus more likely end up in regions that were initially covered by granular material. 
However, this mechanism strongly depends on the details of gravity, impactor- and ejecta-properties such as cohesion (especially for small grains) as well as impact velocities and impact angles. For example, even a particle that rebounds only slowly on Earth might still carry escape velocity of the asteroid after impact which limits the range of relevant impact velocities. So we only consider m/s as relevant in our context though impact speeds in the solar system might easily reach tens of km/s. We expand the work by \citet{Shinbrot2017} here by using well defined particle beds, smaller grains, vacuum and explicitly specifying ejecta energies.

There are a number of studies on granular splashes, i.e. the production of ejecta from an impact into a granular bed \citep{Mitha1986, Werner1990, Colwell1999low, Rioual2000, Rioual2003, Anderson1988}. For example, \citet{Rioual2000} find that 6\,mm impactors into granular beds of the same grain size do not produce ejecta below 6\,m/s impact velocity. Also, \citet{beladjine2007} and \citet{ammi2009three} studied collisions of impacts with 6\,mm grains at tens of m/s and find a linear increase of the ejecta velocity with impact speed. They also find that the fraction of the energy distributed to the ejecta remains constant. It is not known \textit{a priori} how smaller particles would behave at low impact speed and what effect low gravity has on the ejection process. \citet{pacheco2019} study the impact of nonspherical projectiles onto a flat granular surface. They find that with increasing impact energy the crater rim becomes circular regardless of the impactor's shape.


\citet{Andreotti2004two} find a linear relation between the impact velocity $v_0$ and the number of ejecta per splash $N$, the so called splash-function $N = \frac{v_0}{a\sqrt{gd}}-1$, with $a$ being a fitting parameter and $g$ and $d$ the gravitational acceleration and the grain diameter, respectively.
Many others like \citet{Anderson1991, Haff1993, Almeida2006, Zhou2006, Huang2017, Tanabe2017}, to name a few, attend to the detailed characterization of the splash-function and its parameters as well as the theoretical modelling of single grains impacting onto a granular bed.

Impact splashes are further important in the context of sand transport on Earth and Mars \citep{Greeley1980, Bagnold1973, Werner1988, Sauermann2001, kroy2002minimal, parteli2007saltation}


Low gravity influences \citep{Brisset2018} 
on the ejection process and cohesion for dust sized grains are not considered here.
However, some scaling should be possible for slow impacts. The aim of this work is to study slow collisions of sand sized grains, analyze rebound and ejecta and scale the results to asteroid gravity to see if the granular nature of the surface itself is important for the distribution of grains after an impact. We are interested in the redistribution of sand sized matter, so we consider impacts on the order of 1\,m/s or less of sand grains to be important. These slow grains are the ones bound to the asteroid, altering the morphology of its surface in secondary collisions. 

In this work we follow the impact energy. Assuming the impacting particle will not break up upon impact it will redistribute and dissipate a part of its kinetic energy. This can be described by the coefficient of restitution $\varepsilon$ which is defined as the ratio of the magnitudes of the impactor's velocity normal to the target after and before the impact. While typical coefficients of restitution for individual silicate grains are on the order of 0.9 \citep{bogdan2019} the effective coefficient of restitution for a collection of particles can be much lower, especially in low gravity \citep{Sack2013}. What happens is that the initial impact will transfer energy (i.e. by agitation) to the particles surrounding the impact location which in turn will inelastically collide with their neighbours leading to a series of inelastic collisions.

In this work we study in more detail how the energy of an impacting particle gets distributed into the granular system under conditions relevant for asteroid surfaces and further. That is how the injected energy gets distributed between the rebounding impactor, the ejecta created, rearrangements of particles and other dissipative effects.

\begin{figure}
	\includegraphics[width=0.5\textwidth]{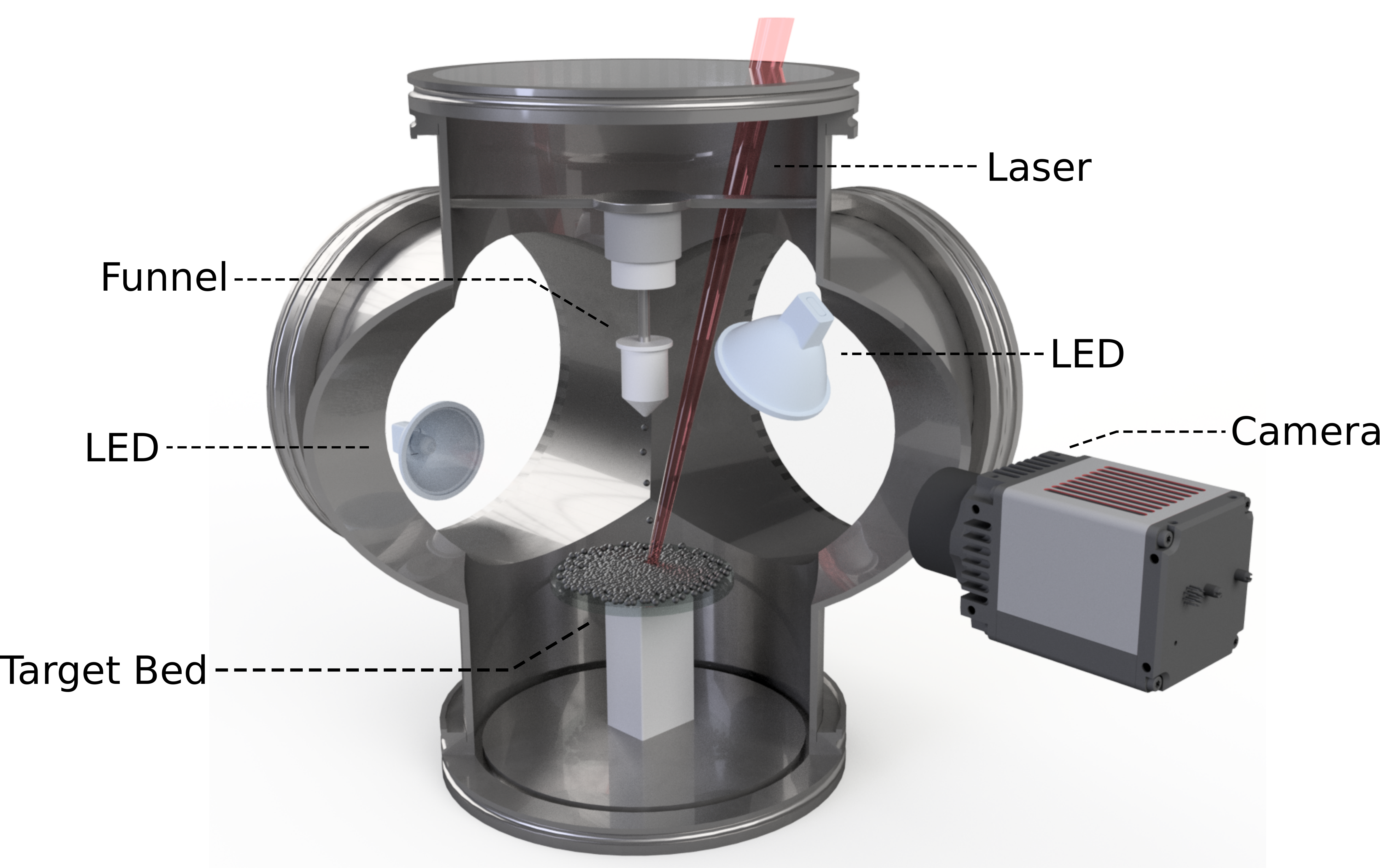}
	\def\svgwidth{0.5\linewidth}
	\caption{\label{fig.setup}Experimental setup. The funnel releases single particles onto the target bed which is illuminated by a laser from above. The impact is recorded with a high-speed camera. LEDs are for general illumination purpose.}
\end{figure}

\section{Experiments}

\begin{figure}
	\includegraphics[width=0.5\textwidth]{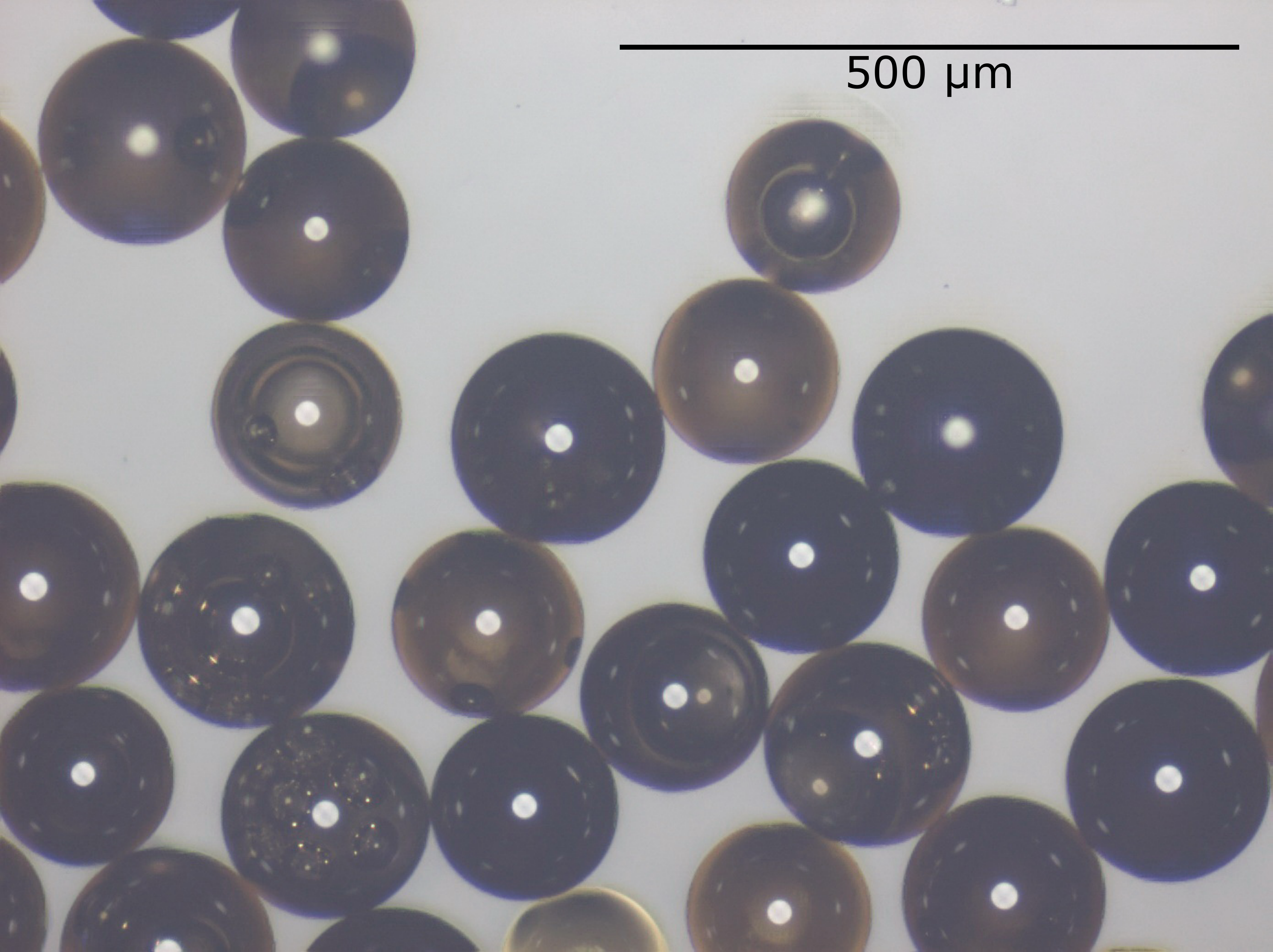}
	\caption{\label{fig.image}Image of the particles used for the experiments. The white dot in the center of the spheres is a specular reflection of the light source illuminating the sample and can be used to track the particle's position.}
	
\end{figure}

To study the granular aspects of slow collisions we designed an experiment where we impact a grain onto a bed of similar sized grains (several 100 000 beads, approx. 2\,cm depth) from low heights under Earth's gravity. The setup consists of a funnel containing a reservoir of particles, which when tapped will release single particles which then fall onto the target surface. Particle impact velocity is set by the mounting height of the funnel and was chosen to be in an interval of 0.25 to 1.19\,m/s. Fig.\,\ref{fig.setup} shows the setup of the experiment and an image of the particles used can be seen in Fig.\,\ref{fig.image}.

\begin{figure}
	\includegraphics[width=0.5\textwidth]{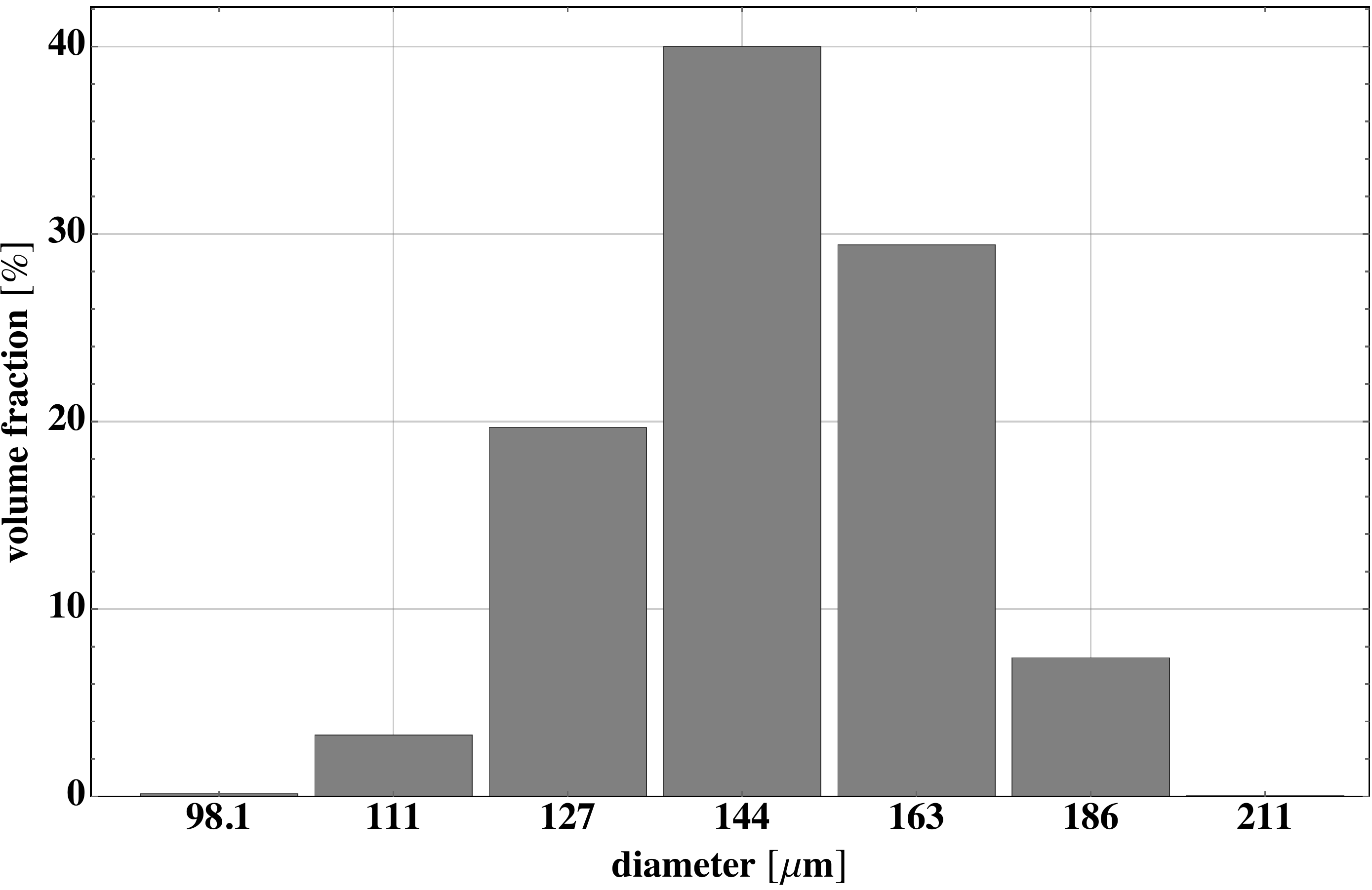}
	\caption{\label{fig.size}Size distribution of the basalt particles. Please note the bars give the volume fraction in the interval around the size noted below the bar. The diameter intervals are not distributed evenly.}
\end{figure}

For this experiment we only consider impacts essentially perpendicular to the target surface, that is we did not vary the impact angle. The setup is evacuated to 6\,mbar to avoid effects of gas drag during movement. At a Reynolds number at normal atmospheric pressure of 8 the grains couple to the gas on the same timescale as the free fall time. It also avoids different behaviours of the granular bed due to interstitial air \citep{pak1995, homan2015}. 
The particles are spherical basalt grains (\emph{Whitehouse Scientific Ltd}) with an average size of 150\,$\rm \mu m$. The detailed size distribution measured by size-dependent light scattering ({\it Malvern Mastersizer 3000}) is shown in Fig.\,\ref{fig.size}. 

As our goal is to quantitatively measure the kinetic energy of moving particles in the system we record their movement. For this we use a high-speed camera recording with a resolution of 187\,px/mm at a frame rate of 3000\,fps and a field of view of about 46 particle diameters. To precisely identify the particles they are illuminated from the top by a laser. This type of illumination gives a clear individual specular reflection on each of the spheres that can be used to trace them using imageJ \citep{Schindelin2012}. With the time resolved particle positions we can then follow the trajectory and velocity of the impactor before and after the impact. The 2D ejecta velocity is calculated from the velocities in two dimensions at the time of the ejection and is based on the height $h_\mathrm{max}$ and the half width $w$ of the parabola 

\begin{equation}
     v_\mathrm{eject}=\sqrt{\frac{w^2 g}{ 2h_\mathrm{max}}+2gh_\mathrm{max}}.
\end{equation} 

\begin{figure}
	\centering
	\includegraphics[width=0.5\textwidth]{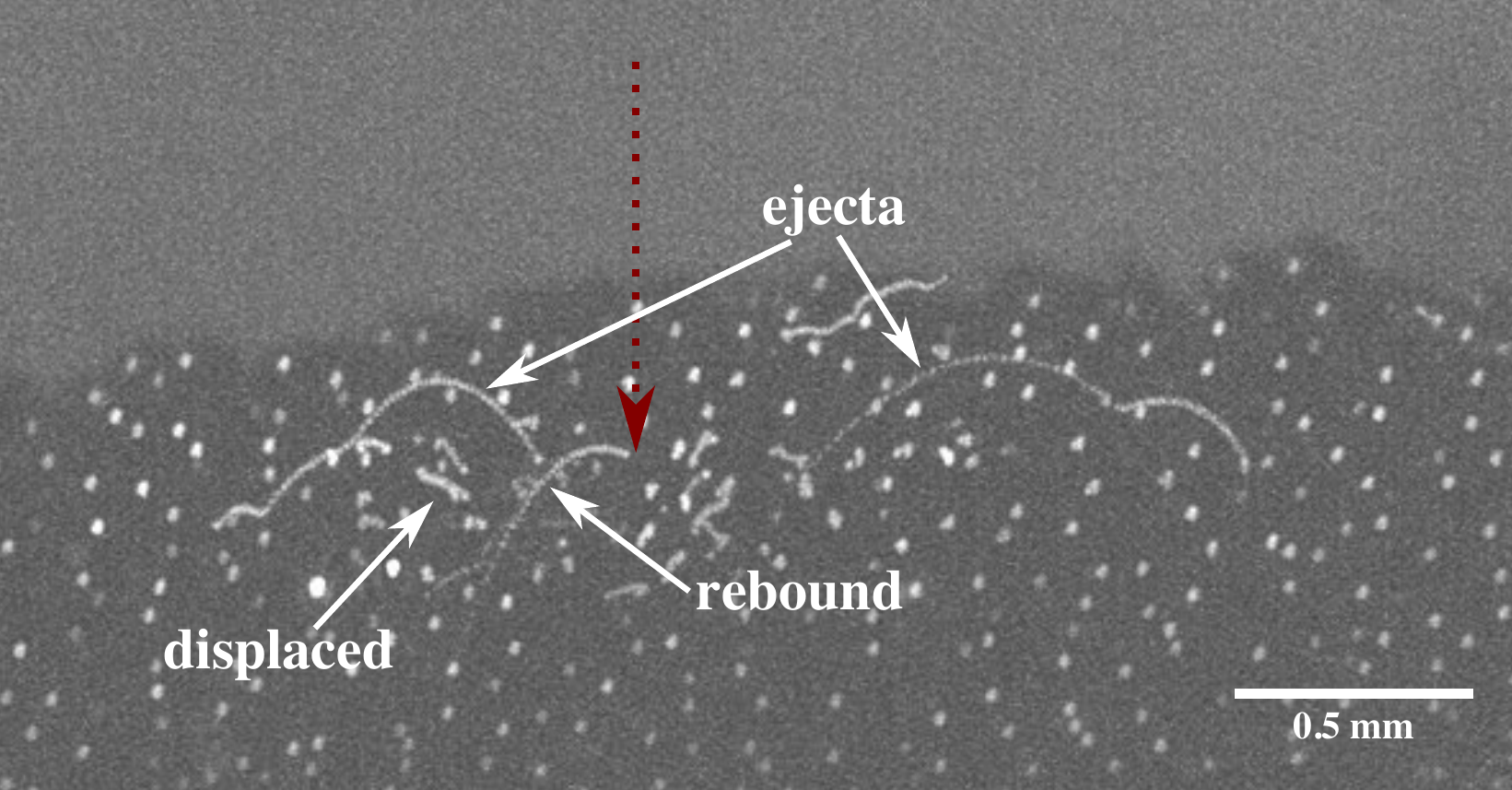}
	\caption{\label{fig.impact} Image of an impact and the consequential reaction to it. The dotted red line represents the trajectory of the impacting grain.}
\end{figure}

Further, we can track the particles on the surface of the target bed. On the target surface we can observe, by eye, three different reactions to the impact: (1) Particles that get ejected (loose contact) from the surface. (2) Particles that do not loose contact with the surface but change their position. (3) Particles that keep their initial position. Fig.\,\ref{fig.impact} is an image showing the particle bed with superimposed particle trajectories. 

To distinguish between cases (1) and (2) we narrow our definition of ejecta to grains that produce trajectories that are clear parabolas. This decision is made by eye. All other particles that do not keep their initial position (i.e. rolling, particles with non-parabolic trajectories) we consider \emph{displaced}. The latter is less constrained as a certain cut-off is necessary for the length of the trajectory. 

\section{Results}

\begin{figure}
	\includegraphics[width=0.5\textwidth]{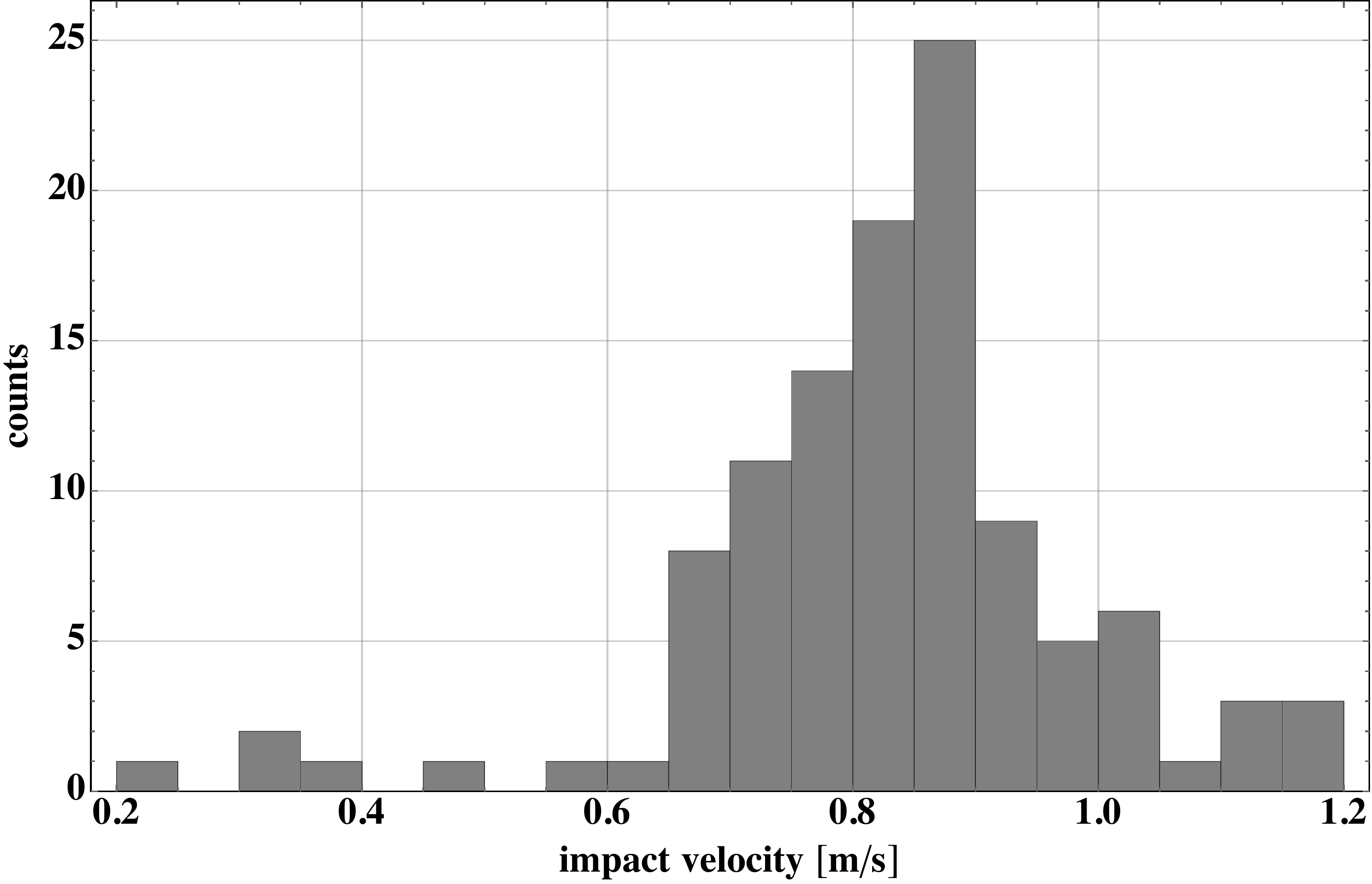}
	\caption{\label{fig.velocities} Distribution of the impact velocities in 111 impact experiments. The mean impact velocity is 0.83\,m/s.} 
\end{figure}

\begin{figure}
	\includegraphics[width=0.5\textwidth]{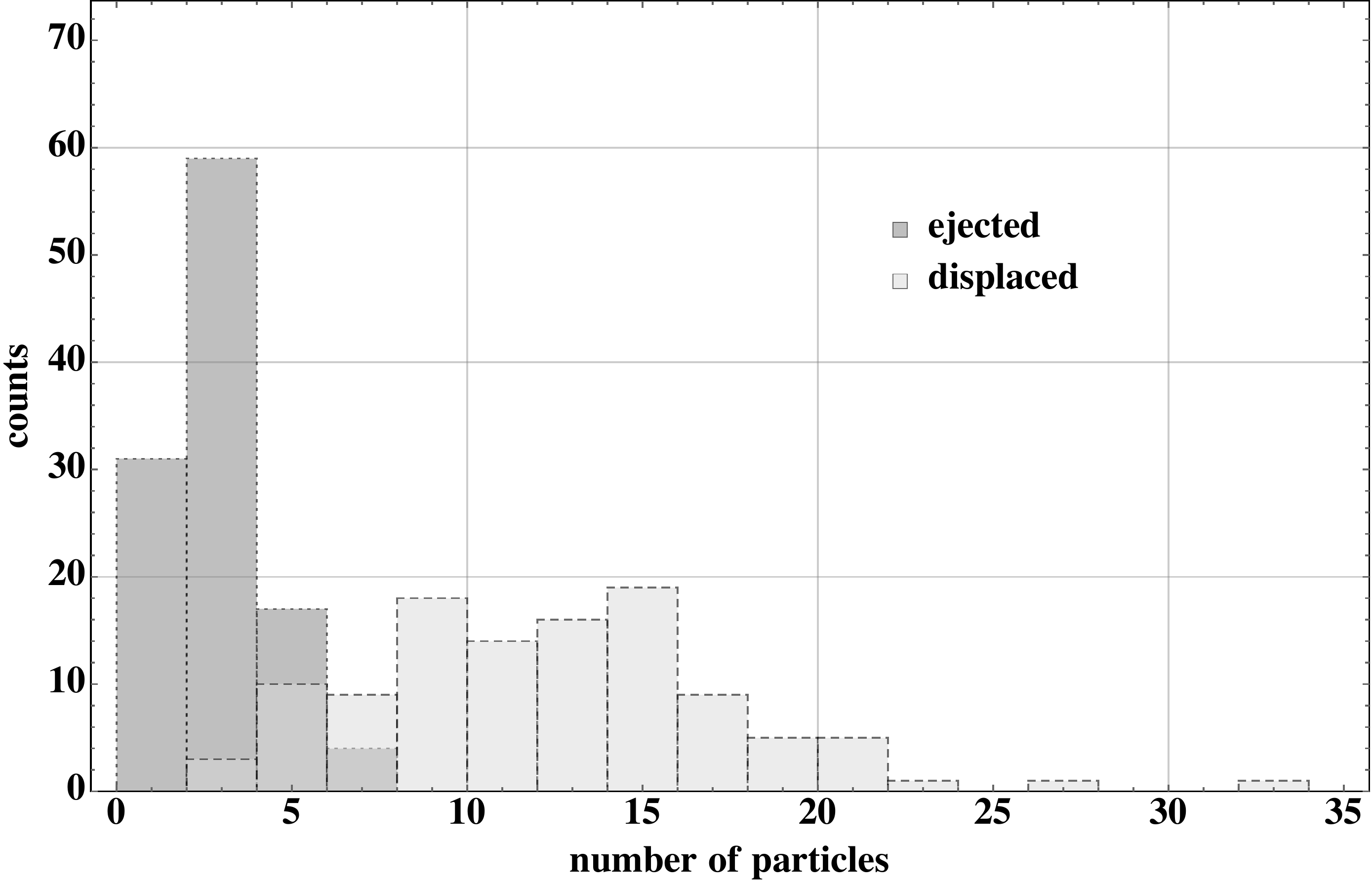}
	\caption{\label{fig.moveit} Distribution of the number of particles affected by the impact. Shown in darker gray with dotted edges is the quantity of how many particles are ejected by a single impacting grain. The lighter gray with dashed edges represents the number of particles that are displaced by a single impacting grain.}
\end{figure}

In total we recorded 111 impact experiments and Fig.\,\ref{fig.velocities} shows the distribution of impact velocities used. 
Fig.\,\ref{fig.moveit} shows that typically a few particles are ejected and somewhat more particles are displaced per impact.

\begin{figure}
	\includegraphics[width=0.5\textwidth]{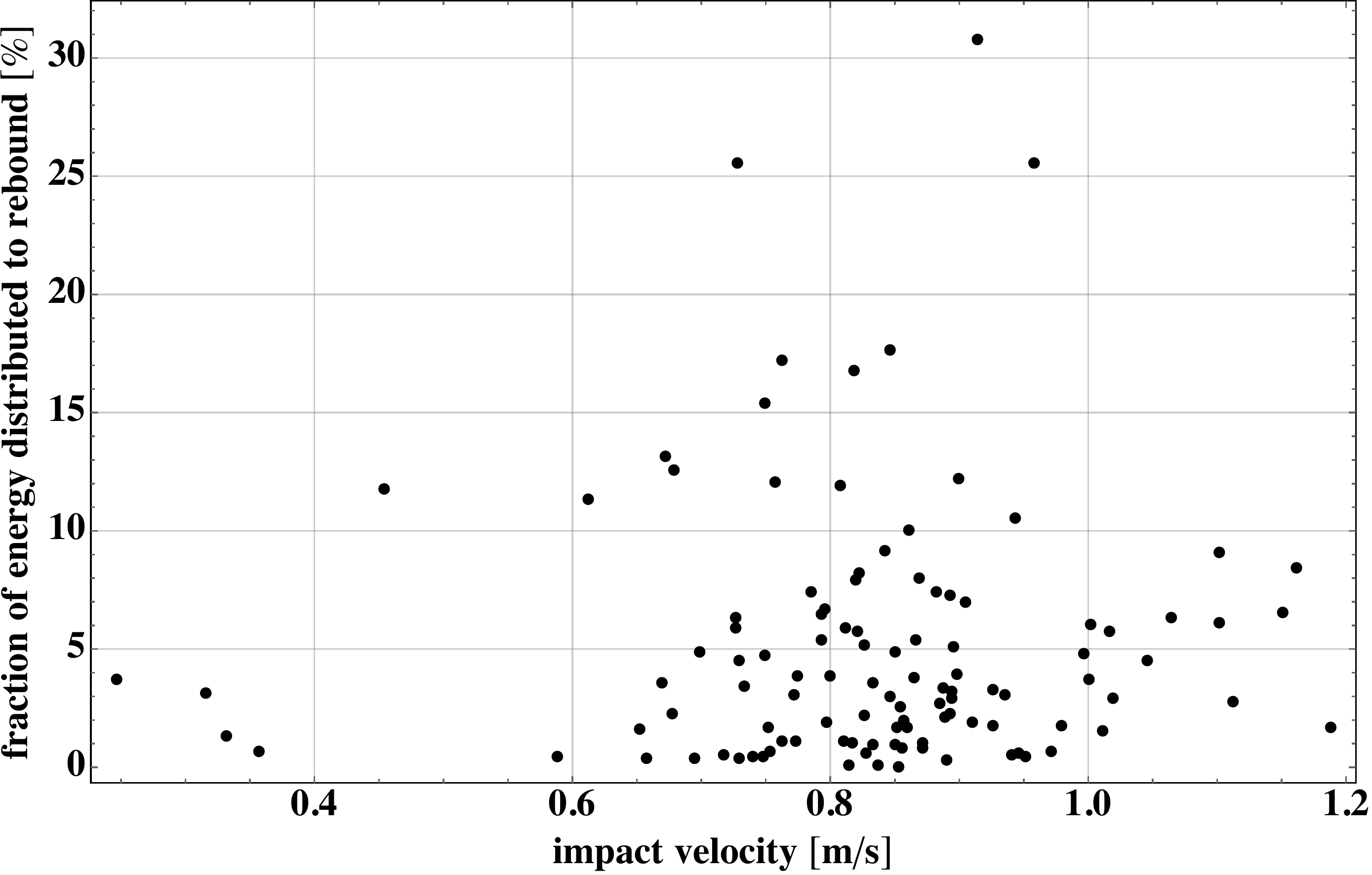}
	\caption{\label{fig.reboundimpact}Fractions of impact energy distributed to the rebounding grain.}
\end{figure}

\begin{figure}
	\includegraphics[width=0.5\textwidth]{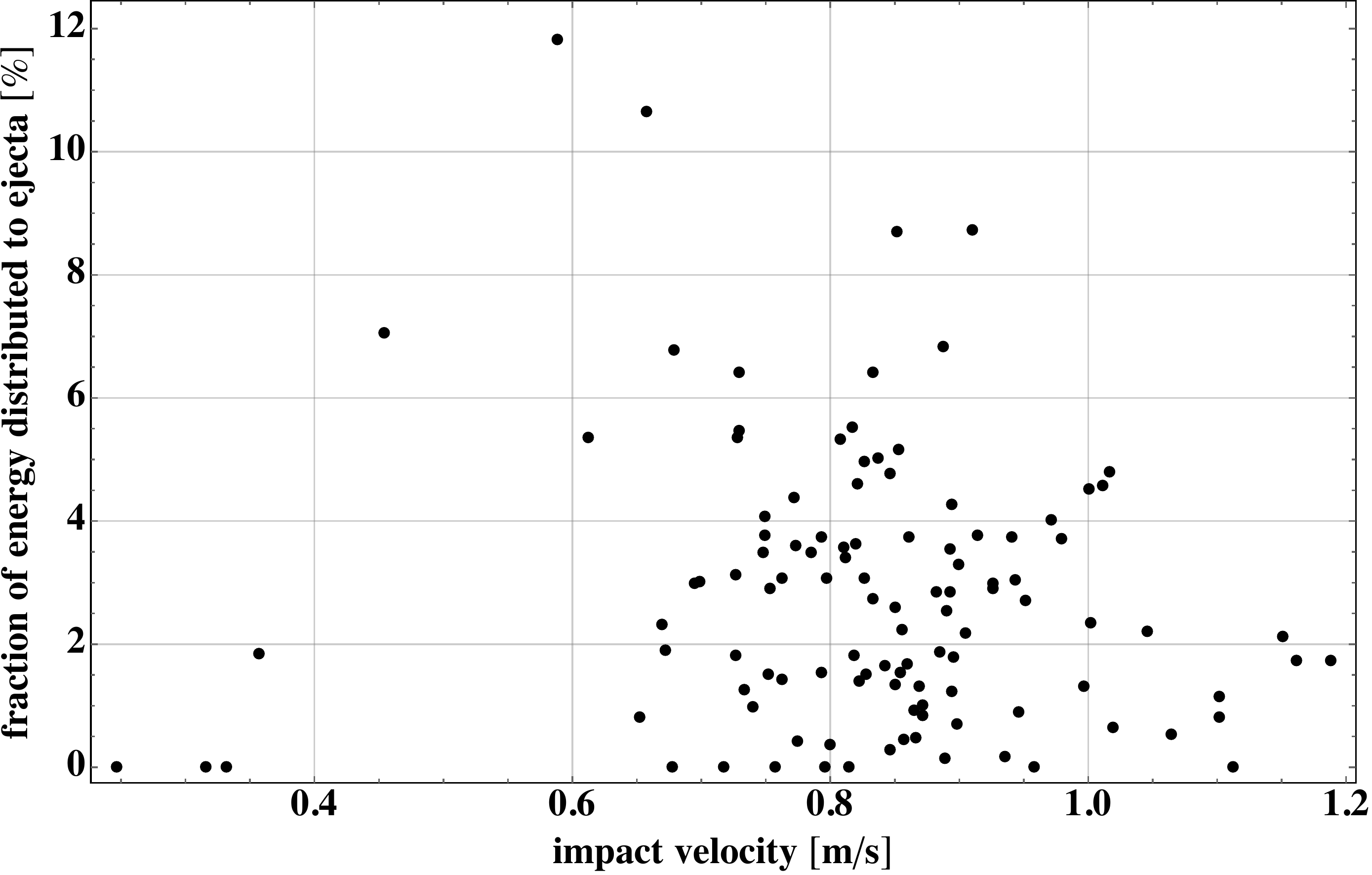}
	\caption{\label{fig.ejectaimpact}Fractions of impact energy distributed to the ejecta.}
\end{figure}

\begin{figure}
	\includegraphics[width=0.5\textwidth]{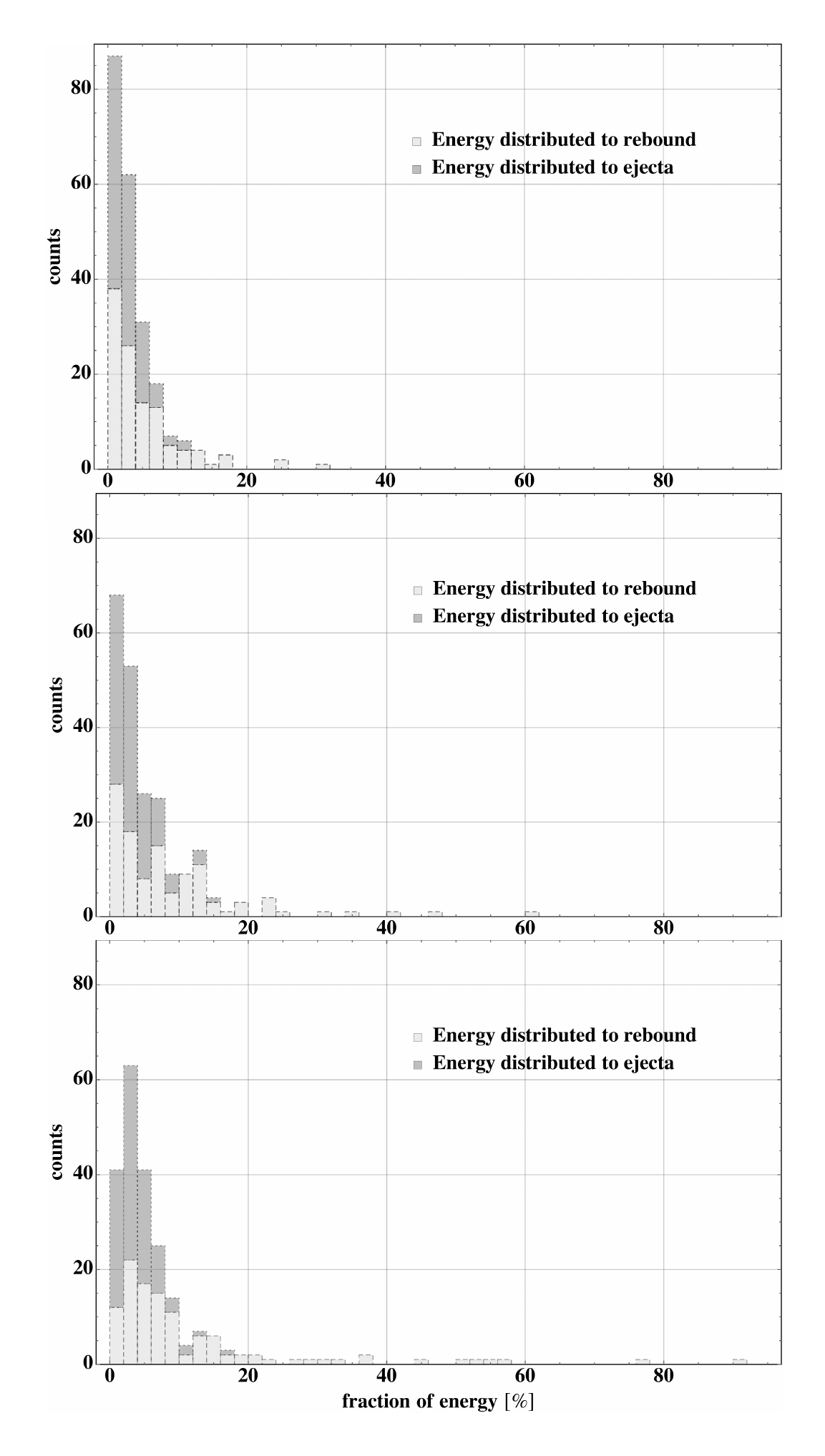}
	\caption{\label{fig.energy2}Fractions of impact energy distributed to the rebounding grain and ejecta. The remaining fraction of the the impact energy is used to rearrange the particle bed or is dissipated otherwise. Top: 2D energies. Center: extreme correction, speeds in both horizontal directions assumed to be the same. Bottom: average horizontal velocity added.}
\end{figure}

From the impact velocities and the known mass of the impactor we calculate the kinetic impact energy for each experiment. We then use the tracked velocity of the rebounding impactor and the ejecta created to calculate their respective kinetic energies. Conservation of energy gives:
\begin{equation} 
E_{ \mathrm{impact}} = E_{\mathrm{rebound}} + E_{\mathrm{ejecta}} + E_{\mathrm{other}} 
\end{equation}
where $E_{\mathrm{other}} $ contains the energy used to displace particles that are not part of the ejecta as well as energy dissipated (i.e. into heat upon impact or through frictional motion in the bed). Fig.\,\ref{fig.reboundimpact} and \ref{fig.ejectaimpact} show the energy distributed to the rebound and the ejecta, respectively. We see no correlation with the impact velocity in agreement to \citet{beladjine2007} and \citet{ammi2009three}. We note though that our velocity range is small. Therefore, we consider all data to follow single distribution functions here. Fig.\,\ref{fig.energy2} shows the distribution of these energies for our experiments. Fig.\,\ref{fig.energy2} top shows the 2D energies measured from the images. Due to the 2D observations the absolute energy is underestimated. To correct for this we considered two different cases. In fig.\,\ref{fig.energy2} center we consider both horizontal directions to be the same. This is an extreme correction as the smallest values get the least corrections and the highest values the largest corrections. In fig.\,\ref{fig.energy2} bottom we added the average horizontal velocity measured to all data. We find that the rebounding impactor carries more energy than the ejecta but the largest portion of the impact energy ends up in $E_{\mathrm{other}}$. From the impactor's perspective the collision is highly dissipative, typically keeping less than 20\,\% of its initial velocity.

From the impact and rebound velocities we can calculate an effective coefficient of restitution for the impactor, which on average in 2D is 0.20 $\pm$ 0.11. However, as can be seen from Figs.\,\ref{fig.reboundimpact}, \ref{fig.ejectaimpact} and \ref{fig.energy2} it is more likely that the energy imparted to impactor and ejecta is small. 

\begin{figure}
	\includegraphics[width=0.5\textwidth]{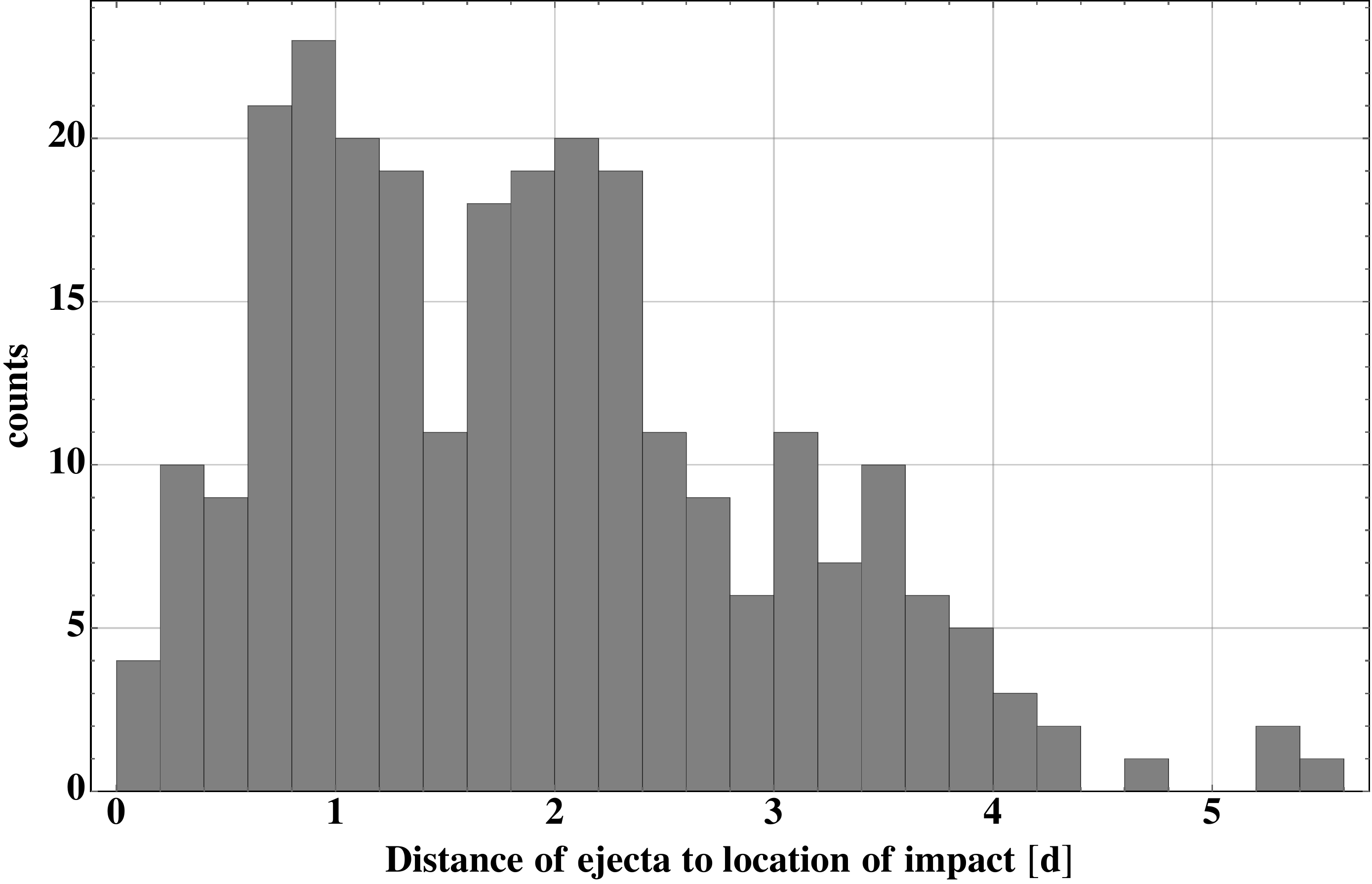}
	\def\svgwidth{0.5\linewidth}
	\caption{\label{fig.distance}Initial location of ejected particles relative to the impact site in units of particle diameter.}
\end{figure}

In our experiment we observe ejecta generation for all initial impactor velocities. Typically an impact event generates between 0 and 5 ejecta particles and displaces about 4 times as many of the spheres. The exact distribution of affected particles can be seen in Fig.\,\ref{fig.moveit}. When following the ejecta trajectories we observe that they are typically generated a few particle diameters away from the impact location, see Fig.\,\ref{fig.distance}.

\section{Application to Asteroids}

Ejecta heights are inversely proportional to gravity. Therefore, on an asteroid with $10^{-4}$\,g ejecta reaching a height of $150\,\rm \mu m$ as measured in our experiment would reach a height of 1.5\,m on the asteroid. Assuming a ballistic arc with similar vertical and horizontal velocities, ejecta would reach a distance of a few meters at maximum. Ejection is therefore still a rather local process for the small particles measured.

The impactor itself rebounds with an average of 5\,\% to 12\,\% of the impact speed depending on the correction. This is on the order of 0.1\,m/s so it can move less than 5\,m on the asteroid considered. This supports the ballistic sorting hypothesis from \cite{Shinbrot2017}.

The results here are only relevant for the studied particle size. In the laboratory as well as on an asteroid inertial forces dominate as the Froude numbers are well beyond 1. The Froude number of the experiment is typically 25 at 1 m/s. On an asteroid it is on the order of 10 000.
In this low gravity environment the particles we study would only travel short distances.
\citet{Rioual2000}, e.g., 
studied larger mm-particles and find typical ejecta heights which, if scaled to the asteroid, are two orders of magnitude larger than our values.
Such particles would travel hundreds of meters which is no longer local and impacts would disperse a granular bed with ejecta of this (large) size ending up everywhere on the asteroid. 

Smaller grains are cohesive and ejection is assumed to be restricted to larger impact speeds. E.g. micrometer grains stick below 1 m/s impact \citep{Poppe2000}.

This leaves a size range of about 0.01 to 1\,mm as impactor and ejecta which would be mobile on an asteroidal surface but if hitting a particle bed grains would not spread far.
A particle bed area of several meters in size or larger will therefore be a sink for these particles. This is especially important in view of rocky parts of an asteroid surface. Typical coefficients of restitution are 0.9 and rebounding impactors will move large distances on the surface. This makes it likely that - after a few collisions - it will eventually hit a particle bed and add to its mass. 
Rock might refer to anything that is significantly larger in mass than the impacting grain, starting from pebble size. 

Particles on asteroids are likely not spherical. How this changes the ejection process in detail is not known. However, as long as the energy distribution will not be dominated by the contact forces the situation should be similar for similar sized particles, irrespective of irregularities. So we consider that our results still hold for irregular grains, though this should certainly be verified in future experiments.

\section{Conclusion}

We designed and conducted an experiment to study slow impacts of small particles into granular beds of the same material as the impactor. We showed that in these collisions into granular material most of the incoming energy is dissipated and only a small fraction of the energy remains in the rebounding impactor as well as the generated ejecta. We also see that grains are not ejected at the point of impact but typically a few particle diameters away. In total, the slowly rebounding impactor and the slow ejecta generated confine the effect of the impact to a small area. In that sense our data supports the idea by \cite{Shinbrot2017} that small grains gather among themselves and leave larger outcrops free of grains. However, we can put constraints on the size sorting effect on asteroids. The observed splash distances give a lower limit for the size of the initial granular bed needed to support a ballistic sorting regime. Asteroids are observed to have size-sorted terrains. We showed here that this mechanism especially holds for the low gravity environment of asteroids. Beyond that also the transport of grains on surfaces of other low gravity planetary bodies, like Pluto or Titan, might be influenced by such slow splashes \citep{lorenz2006sand, telfer2018dunes}.

\section*{Acknowledgements}
This project was funded by DFG grant WU 321/18-1 and supported by the DFG under grant WU 321/14-1.
This work was also supported by the DLR Space Administration with funds provided by the Federal Ministry for Economic Affairs and Energy (BMWi) based on a decision of the German Federal Parliament under grant number 50WM1943. 

\newpage
\bibliographystyle{aa}
\bibliography{bib}

\end{document}